\documentclass[10pt,twocolumn,letterpaper]{article}

\usepackage[pagenumbers]{cvpr} 

\usepackage[dvipsnames]{xcolor}

\usepackage[accsupp]{axessibility} 

\usepackage{xcolor}         
\usepackage{adjustbox}
\usepackage{amsmath}
\usepackage{amsthm}
\usepackage{amsfonts,amssymb} 
\usepackage{graphicx}
\usepackage{algorithm}
\usepackage{algorithmic}
\usepackage{setspace}
\usepackage{multirow}
\usepackage{pifont}
\usepackage{bbding}
\usepackage{color}
\usepackage{wrapfig}
\usepackage{bm}
\usepackage{colortbl}

\definecolor{cvprblue}{rgb}{0.21,0.49,0.74}
\usepackage[pagebackref,breaklinks,colorlinks,citecolor=cvprblue]{hyperref}

\def\paperID{8063} 
\def\confName{CVPR}
\def\confYear{2024}

\newcommand{\F}		{\mathcal{F}}
\newcommand{\C}     {\mathbb{C}}

\newcommand{\N}     {\mathcal{N}}
\newcommand{\herm}{\mathsf{H}}
\newcommand{\y}     {\bm{y}}
\newcommand{\m}     {\bm{m}}
\newcommand{\p}     {\bm{p}}
\newcommand{\e}     {\bm{e}}
\newcommand{\x}     {\bm{x}}
\newcommand{\z}     {\bm{z}}

\DeclareMathOperator*{\argmin}{argmin}

\def\ie{\emph{i.e., }}
\def\eg{\emph{e.g., }}

\title{Progressive Divide-and-Conquer via Subsampling Decomposition for Accelerated MRI}

\author{Chong Wang, Lanqing Guo, Yufei Wang, Hao Cheng, Yi Yu, Bihan Wen\thanks{Bihan Wen is the corresponding author.}\\
Nanyang Technological University, Singapore\\
{\tt\small 	\{wang1711, lanqing001, yufei001, HAO006, yuyi0010, bihan.wen\}@ntu.edu.sg}
}

\begin{document}
\newtheorem{definition}{Definition}

\maketitle
\begin{abstract}
Deep unfolding networks (DUN) have emerged as a popular iterative framework for accelerated magnetic resonance imaging (MRI) reconstruction.
However, conventional DUN aims to reconstruct all the missing information within the entire null space in each iteration. 
Thus it could be challenging when dealing with highly ill-posed degradation, usually leading to unsatisfactory reconstruction.
In this work, we propose a Progressive Divide-And-Conquer (PDAC) strategy, aiming to break down the subsampling process in the actual severe degradation and thus perform reconstruction sequentially.
Starting from decomposing the original maximum-a-posteriori problem of accelerated MRI, we present a rigorous derivation of the proposed PDAC framework, which could be further unfolded into an end-to-end trainable network.
Specifically, each iterative stage in PDAC focuses on recovering a distinct moderate degradation according to the decomposition.
Furthermore, as part of the PDAC iteration, such decomposition is adaptively learned as an auxiliary task through a degradation predictor which provides an estimation of the decomposed sampling mask.
Following this prediction, the sampling mask is further integrated via a severity conditioning module to ensure awareness of the degradation severity at each stage.
Extensive experiments demonstrate that our proposed method achieves superior performance on the publicly available fastMRI and Stanford2D FSE datasets in both multi-coil and single-coil settings.
Code is available at \url{https://github.com/ChongWang1024/PDAC}.
\end{abstract}

\vspace{-0.8cm}

\begin{figure} 
  \centering 
    \includegraphics[width=\linewidth]{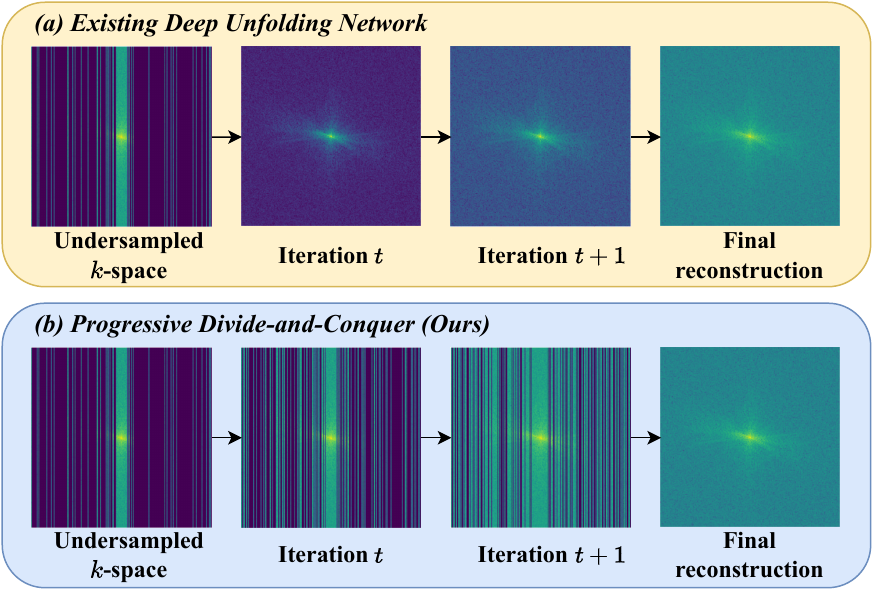}
    \vspace{-6.5mm}

    \caption{In each iteration: (a) Existing deep unfolding networks recover the information in the entire null space;
    (b) Our proposed method decomposes the entire null space and selectively retrieves information within specific segments of the null space, progressing from those that are easier to recover to more challenging ones.}
    \vspace{-4.5mm}
    \label{fig_intro}
  \end{figure} 

\section{Introduction}
\label{sec:intro}
Magnetic resonance imaging (MRI) is a widely used non-invasive imaging technique that provides detailed visualization of anatomical structures for medical diagnosis.
MRI scanners sequentially collect data in the frequency domain, commonly referred to as $k$-space, from which MR images are reconstructed.
{One inherent limitation of MRI is the protracted data acquisition time, which can cause patient discomfort and result in artifacts in the reconstructed images due to physiological movements during the acquisition.}
Thus accelerating the data acquisition process has been a primary focus in the MRI community.
{
Compressed sensing (CS)~\cite{donoho2006compressed, lustig2007sparse} is a common approach to accelerate the measurement acquisition process of MRI. 
These approaches employ undersampled data in $k$-space and exploit sparsity and incoherence to achieve accurate reconstructions. 
}
Specifically, traditional CS techniques necessitate that the underlying MR images exhibit sufficient sparsity in certain transform domains or learned dictionaries~\cite{ravishankar2015efficient}. 
However, these assumptions can be overly restrictive for reconstructing MR images subjected to various degradations, 
limiting the performance of these traditional approaches.

Recent advancements in deep learning have demonstrated superior performance in MRI reconstruction by leveraging a large collection of training data~\cite{bhadra2021hallucinations, schlemper2017deep, cheng2019model, zheng2019cascaded, jalal2021robust}.
Deep unfolding networks (DUN)~\cite{zhang2020deep, zheng2022unfolded, admmnet}, in particular, have emerged as a reliable framework for MRI reconstruction by unrolling an iterative optimization algorithm into an end-to-end trainable network.
This approach establishes a potent deep framework that integrates explicit constraints based on the degradation model, offering a crucial foundation for effectively addressing ill-posed reconstruction problems.
However, the efficacy of this constraint may be compromised, especially when solving highly ill-posed reconstruction problems, as it solely regularizes the limited range space of the sensing matrix.
Figure~\ref{fig_intro}~(a) shows one example in which a conventional DUN aims to reconstruct the full samples from the highly undersampled measurements in $k$-space in each iteration. Consequently, the reconstruction quality may degrade due to cumulative errors.

In this work, we present a novel iterative strategy, namely Progressive Divide-and-Conquer (PDAC), by decomposing the subsampling process in the actual severe degradation and thus performing reconstruction sequentially.
Specifically, we initiate by reformulating the original maximum-a-posteriori problem of accelerated MRI in a way that the original severe problem can be progressively decoupled into a series of moderate corruptions.
Accordingly, we present a rigorous derivation of the reconstruction process of PDAC, which is further unfolded into an end-to-end trainable network.
With the merit of degradation decomposition, each iteration in PDAC focuses solely on retrieving the information within the specific parts of null space, specifically targeting a particular moderate degradation according to the decomposition, as shown in Figure~\ref{fig_intro}~(b).
Besides, such decomposition is adaptively learned as an auxiliary task throughout the PDAC iterations.
Specifically, a degradation predictor is adopted to predict the subsampling mask that characterizes the decomposed degradation process. 
This learning-to-decompose mechanism serves to guide the preservation of specific reconstructed information, concurrently discarding inaccuracies.
Consequently, each intermediate reconstruction is characterized by a specific degradation severity, as indicated by the predicted subsampling mask.
Thus we further introduce an embedding module to guarantee awareness of the degradation severity at each iterative stage.
Experimental results show that the proposed method can achieve superior performance consistently in both single-coil and multi-coil MRI reconstruction on the fastMRI and Stanford2D FSE datasets.

The main contributions of this work are as follows:
\begin{itemize}[leftmargin=6mm]
    \item {We propose the Progressive Divide-and-Conquer (PDAC) method for accelerated MRI, a novel iterative strategy that decomposes the subsampling process of severe degradation into a series of moderate ones and then performs reconstruction sequentially. 
    In PDAC, each iteration is dedicated to recovering a specific decomposed degradation.
    }
    \item We unfold the PDAC iterations into an end-to-end trainable network, simultaneously incorporating the learning of degradation decomposition as an auxiliary task.
    To guarantee awareness of decomposed degradation at each iteration, we introduce a severity embedding module, facilitating the integration of the decomposed subsampling mask into our network.
    \item Extensive experimental results on the fastMRI and Stanford2D FSE datasets in both single-coil and multi-coil MRI reconstruction show that the proposed PDAC achieves superior performance compared to the state-of-the-art.
    
\end{itemize}

\section{Related Work}
\label{sec:relatedwork}
\subsection{Accelerated MRI}
The success of reconstructing images in CS-MRI relies on incorporating an additional regularizer capturing assumed image priors due to its ill-posed nature. 
Traditional methods commonly employ handcrafted priors such as total variation~\cite{tv_mri} and low-rankness~\cite{low_rank} to regulate the reconstruction results. In addition to these approaches, dictionary learning~\cite{dictionary} stands out as a more flexible method for MRI reconstruction, directly learning an adaptive sparse representation from the data.
More recently, deep learning-based methods~\cite{muckley2021results, xin2023fill, wang2023deep, nitski2020cdf, alghallabi2023accelerated, fabian2022humus} have achieved remarkable performance in MRI reconstruction.
CDF-Net~\cite{nitski2020cdf} introduces a cross-domain fusion network for MRI reconstruction by exploiting the relationship in both frequency and spatial domains.
Specifically, model-based learning methods have emerged as a popular paradigm for solving MRI reconstruction.
For example, ADMM-Net~\cite{admmnet} attempts to learn the parameters of the data flow in the alternating direction method of multipliers (ADMM) algorithm and has demonstrated superior results for MRI reconstruction. 

\subsection{Deep Unfolding Networks}
By unfolding deep networks that embody the steps from existing model-based iterative algorithms and optimizing in an end-to-end manner, DUN has demonstrated promising performance on various inverse problems, such as super-resolution~\cite{zhang2020deep, zheng2022unfolded} and deblurring~\cite{kong2021deep, mou2022deep}. 
DUN also becomes a predominant approach for solving MRI reconstruction~\cite{admmnet, arvinte2021deep, yang2020model, jun2021joint} that ensures the consistency of the predictions with respect to the degradation model.
The earliest unrolling method can be traced back to LISTA~\cite{gregor2010learning} where the authors propose to unfold the iterative shrinkage thresholding algorithm (ISTA)~\cite{ista} for sparse coding.
MoDL~\cite{MoDL} introduces a model-based learning framework by unfolding a weight-sharing network with several iterations, thus offering appealing performance with less computational cost.
\cite{zhou2020dudornet} exploits dual domain priors on both image space and $k$-space to simultaneously constrain the MRI reconstruction results via a customized dual domain recurrent network.
However, existing unfolding methods aim to recover the information in the entire null space at each stage, which can lead to inaccurate reconstructions, particularly when dealing with highly degraded problems.
In this paper, we introduce a novel unfolded paradigm that decomposes the degradation and progressively retrieves specific segments of the null space, targeting a particular decomposed degradation.

\section{Preliminary}
\subsection{MRI Acquisition}
\label{sec:problem}
Compressed sensing MRI accelerates the data acquisition process by collecting less data than the Nyquist rate in the $k$-space.
The MR image is reconstructed by maximum a posteriori (MAP) problem, involving the data fidelity term and regularizer $\mathit{\Psi}(\cdot)$, which is
\small
\begin{equation}\label{eq_prob_P0}
    \min_{\x}   \lVert \y - A \x \rVert^2_2 + \lambda \mathit{\Psi}(\x),
\end{equation}
\normalsize
where $\x\in \C^n$ denotes the underlying spatial ground truth image, $\y \in \C^n$ is the zero-filled undersampled measurements, $A \in \mathbb{C}^{n \times n}$ denotes the degradation process and $\lambda$ is the weighting parameter that controls the penalty strength of the regularizer.

\noindent \textbf{Single-coil MRI.}
In single-coil MRI, the degradation process corresponds to subsampling in the $k$-space
\begin{equation}\small\label{eq_forward_single}
    \y = A \x + \bm{\epsilon} = D \F \x + \bm{\epsilon},
\end{equation}
where $D=\text{diag}(\bm{d}) \in \mathbb{R}^{n\times n},\  \bm{d} \in \{0, 1\}^n $ is a diagonal sampling matrix characterizing the sampled location in $k$-space, $\F \in \C^{n \times n}$ denotes the two-dimensional Fourier transform and $\bm{\epsilon} \in \C^n$ is the measurement noise during the acquisition process.
The ill-posedness stems from the underdetermination of $D$ (\ie $\text{rank}(D)=m \ll n$).

\noindent \textbf{Multi-coil MRI.}
In the multi-coil MR system with $C$ coils, the undersampled measurement obtained by the $c$th coil can be formulated as 
\begin{equation}\small\label{eq_forward_multi}
    \y_c =  D \F S_c \x + \bm{\epsilon}_c, \quad c=1, \dots C,
\end{equation}
where the coil-sensitivity map $S_c\in\C^{n\times n}$ is a matrix encoding the sensitivity of the $c$-th coil.
The coil-sensitivity map is usually normalized such that $\sum_{c=1}^C S_c^{\herm} S_c = I$.


\subsection{Half Quadratic Splitting}
Optimization problem~\eqref{eq_prob_P0} can be solved using variable splitting techniques such as half-quadratic splitting~\cite{hqs} by introducing an auxiliary variable $\z$, that reformulates~\eqref{eq_prob_P0} as 
\begin{equation}\small
        \min_{\z, \x}   \lVert  \y - A \z \rVert^2_2 +  {\mu}  \lVert \z - \x \rVert^2_2 + \lambda \mathit{\Psi}(\x),
\end{equation}
where $\mu$ is the penalty parameter of the Lagrangian term.
Then it can be solved using alternating minimization between two sub-problems as 
\small
\begin{align}\label{eq_hqs_z}
    \z_t &= \argmin_{\z}   \lVert  \y - A \z \rVert^2_2 +  {\mu}  \lVert \z - \x_{t-1} \rVert^2_2,\\ \label{eq_hqs_x}
    \x_t &= \argmin_{\x}  {\mu}  \lVert \x - \z_t \rVert^2_2 + \lambda \mathit{\Psi}(\x). 
\end{align}
\normalsize
Note that~\eqref{eq_hqs_z} is a least-squares problem with a quadratic penalty term which has a closed-form solution, commonly dubbed as the data consistency step.
The update process of $\x_t$ corresponds to a Gaussian denoising problem that can be solved using any denoiser.
By adopting a deep network as the regularizer in~\eqref{eq_hqs_x}, the above iterative steps can be unfolded into an end-to-end trainable framework that is alternatively optimized between~\eqref{eq_hqs_z} and \eqref{eq_hqs_x}.

\begin{figure*} 
  \centering 
    \includegraphics[width=0.9\linewidth]{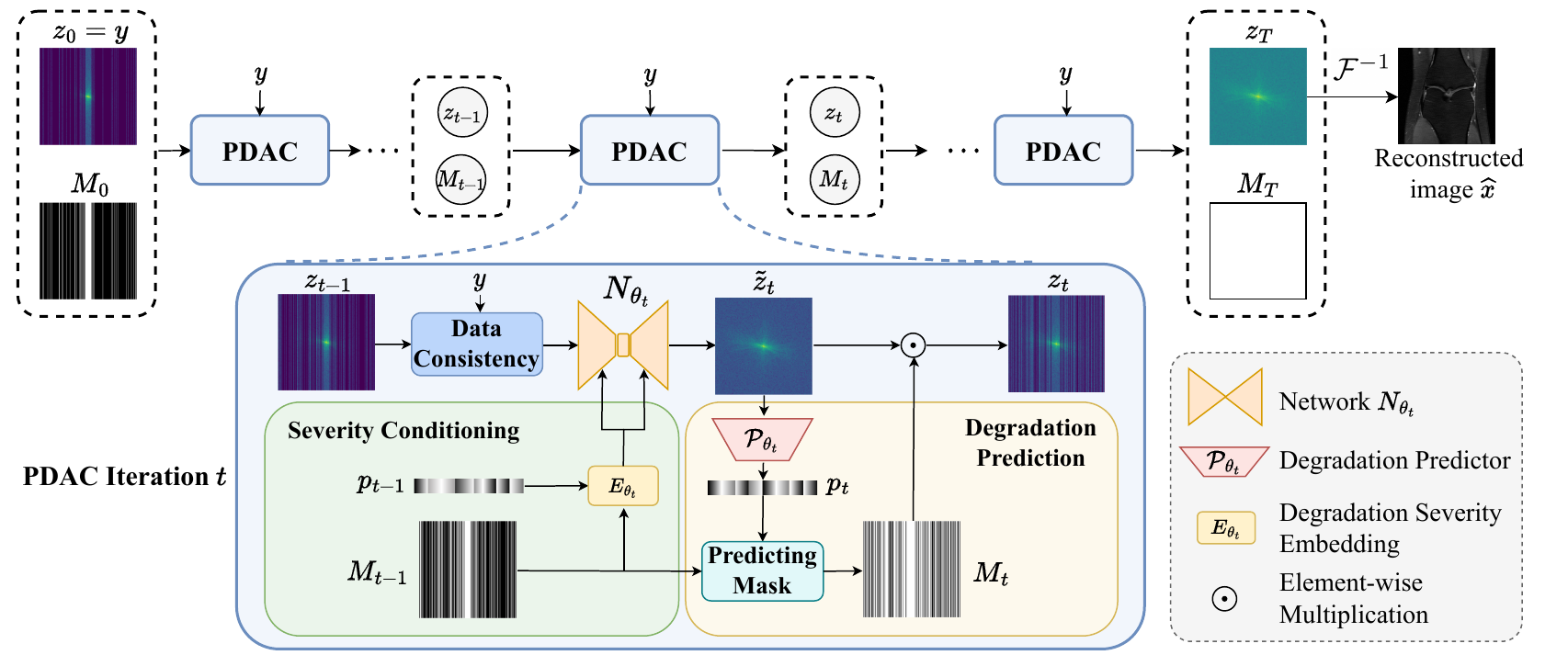}
    \vspace{-2mm}

    \caption{Illustration of the proposed \textbf{\textit{progressive divide-and-conquer (PDAC)}} framework, where the iterative process is detailed in Algorithm~\ref{alg1}.
    Each iteration consists of data consistency, network reconstruction, and degradation using $M_t$.
    Besides, the network learns the decomposed degradation, characterized by $M_t$, as an auxiliary task along iterations via
    (1) Degradation Prediction: we adaptive learn a decomposed sampling mask $M_t$ which indicates the frequency components to preserve in $\tilde{\z}_t$;
    (2) Severity Conditioning: we adopt a severity embedding module $E_{\theta_t}$ to guarantee awareness of the degradation pattern in $M_{t-1}$.}
    \label{fig_framewwork}
    \vspace{-2.0mm}
  \end{figure*} 

\section{Method}
\label{sec:method}
Recovering the degradation $A$ directly poses significant challenges, especially when $m \ll n$.
To this end, we introduce the progressive divide-and-conquer (PDAC) algorithm, which initiates by breaking down a severe degradation into a series of moderate corruptions and recovering each decomposed degradation sequentially.
Subsequently, we unfold the PDAC iterations into an end-to-end trainable network while simultaneously learning the degradation decomposition as an auxiliary task during the iteration.
The overall framework of PDAC is illustrated in Figure~\ref{fig_framewwork}.
\subsection{Decomposing the Subsampling}\label{subsec_decompose}
The zero-filled degradation $A$ can be decoupled into two parts as $A=D\F$.
Specifically, $D$ indicates the sampling matrix where the ill-posedness comes from since $D$ is underdetermined.
We can decompose such severe degradation caused by subsampling $D$ into a product of multiple moderate degradations as
\begin{equation}\small\label{eq_decompose}
\begin{split}
    D = D_0 \cdots D_T= D_0 \cdots D_{t-1} \bar{D}_t
\end{split}
\end{equation}
where $D_t=\text{diag}(\bm{d_t}), \ \bm{d_t} \in \{0, 1\}^n$ is the decomposed diagonal sampling matrix and {\small{$\bar{D}_t = (\prod_{i=t}^T D_i)$}}. 
Specifically, {\small{$\text{rank}(\bar{D}_t)<\text{rank}(\bar{D}_{t+1})$, $\bar{D}_T = I$}}.
Let $A_t= \bar{D}_t \F$ and we have $A_0 = A$, $A_T = \F$.
Since the ill-posed nature of the problem only stems from the sampling matrix $D$, we now present a formal definition of the degradation severity of $A_t$ according to the range space of its corresponding $\bar{D}_t$.

\begin{definition}\label{definition}
\textbf{\textit{(Degradation severity)}}. The degradation $A_t$ is more severe when the range space of the sampling matrix $\bar{D}_t^\top$ is a subspace of that in $A_{t'}$.
According to the decomposition in~\eqref{eq_decompose}, we have 
\begin{equation}\small
\begin{split}
        \mathrm{R}(\bar{D}_t^{\top}) \subset \mathrm{R}(\bar{D}_{t'}^{\top}) \subset \mathrm{R}(\bar{D}_T^{\top}) = \mathbb{R}^n, \quad  \forall \ t < t' < T, \\
         \mathrm{R}(A_t^{\herm}) \subset \mathrm{R}(A_{t'}^{\herm}) \subset \mathrm{R}(A_T^{\herm}) = \mathbb{C}^n, \quad  \forall \ t < t' < T.
\end{split}
\end{equation}
\end{definition}
This decomposition provides a means of disentangling severe corruption into a series of moderate ones.
Following this, we will introduce a progressive divide-and-conquer strategy, addressing the original severe degradation problem iteratively through degradation decomposition.

\subsection{Progressive Divide-and-Conquer}
Solving~\eqref{eq_prob_P0} with a single regularizer can be challenging when the measurement $\y$ is significantly degraded.
Alternatively, we propose to find a solution by introducing a series of regularizers according to the decomposed degradation defined in~\eqref{eq_decompose}, and the problem can be reformulated as 
\small
\begin{equation}\label{eq_prob_P1}
    \min_{\x}   \lVert \y - A_0 \x \rVert^2_2  + \sum_{t=1}^T \lambda_t \mathit{\Psi}_t(A_t \x), 
\end{equation}
\normalsize
where $\{\mathit{\Psi}_t(A_t \x)\}_{t=1}^T$ are the regularizers encapsulating the prior knowledge on the distribution of less degraded measurements $A_t \x$ and $\lambda_t$ is the weighting parameter accordingly.
\eqref{eq_prob_P1} can be solved by adopting variable splitting algorithms such as half-quadratic splitting (HQS)~\cite{hqs}.
Specifically, we can write the augmented Lagrangian form of~\eqref{eq_prob_P1} by introducing an auxiliary variable $\z_1 = A_1 \x$ as:
\begin{equation}\small
\begin{split}
    \min_{\z_1, \x}   \lVert  \y - &D_0 \z_1 \rVert^2_2 +  {\mu_1}  \lVert \z_1 - A_1 \x \rVert^2_2 + \lambda_1 \mathit{\Psi}_1(\z_1)  \\ 
    &+ \sum_{t=2}^T \lambda_t \mathit{\Psi}_t(A_t \x) , \quad \ \text{s.t.} \ \z_1\in \mathrm{R}(A^{\herm}_1).
\end{split}
\end{equation}
The constraint implies that the intermediate measurement $\z_1$ should lie in the range space of decompose degradation $A_1$, as indicated by the Lagrangian term $\lVert \z_1 - A_1 \x \rVert^2_2$.
Then it can be decoupled into two sub-problems as 
\small
\begin{align}
\begin{split}
    \z_1 = \argmin_{\z_1}   &\lVert \y  - D_0 \z_1 \rVert^2_2 +  {\mu_1}   \lVert \z_1 - A_1 \x \rVert^2_2 \\
    &\qquad \qquad+ \lambda_1 \mathit{\Psi}_1(\z_1), \quad \ \text{s.t.} \ \z_1\in \mathrm{R}(A^{\herm}_1)\\ 
\end{split} \label{eq_update_z1}  \\
    &\min_{\x}  {\mu_1}  \lVert \z_1 - A_1 \x \rVert^2_2  + \sum_{t=2}^T \lambda_t \mathit{\Psi}_t(A_t \x).  \label{eq_update_x}
\end{align}
\normalsize
Problem~\eqref{eq_prob_P1} characterized by the degradation $A_0$ is now converted to a less severe problem~\eqref{eq_update_x} characterized by $A_1$ via introducing an intermediate measurement $\z_1$.
Solving the sub-problem in~\eqref{eq_update_z1} can be interpreted as recovering the information loss during the decomposed subsampling $D_0$, thus transforming the optimization problem into a less severe one.

Unfortunately, directly solving~\eqref{eq_update_x} could still be challenging.
One can perform a similar variable splitting by introducing variable $\z_2 = A_2 \x$ that splits~\eqref{eq_update_x} into another two sub-problems via updating $\z_2 $ and $\x$. 
In this case, the overall optimization problem~\eqref{eq_prob_P1} could be approximated via solving three variables, $\z_1, \z_2$ and $\x$, sequentially.
Towards this end, such a variable splitting trick can be performed progressively by introducing a series of auxiliary variables $\{\z_t = A_t \x\}_{t=1}^T$, leading to solving sequential sub-problems $\{\z_1, \dots,\z_t\}$ and $\x$ as
\small
\begin{align}
\begin{split} \label{eq_update_zt}
    \z_\tau = &\argmin_{\z_\tau\in \mathrm{R}(A^{\herm}_\tau)}  \ {\mu_{\tau-1}}  \lVert \z_{\tau-1}  - D_{\tau-1} \z_\tau \rVert^2_2 \\
    & +  {\mu_\tau}   \lVert \z_\tau - A_\tau \x \rVert^2_2 + \lambda_\tau \mathit{\Psi}_\tau(\z_\tau), \quad \forall \ \tau \in [1, t],\\ 
\end{split}\\
    &\min_{\x}  {\mu_t}  \lVert \z_t - A_t \x \rVert^2_2 + \sum_{i=t+1}^T \lambda_i \mathit{\Psi}_i(A_i \x). 
\end{align}
\normalsize

Note that the final estimation $\x$ only updates once, after we have sequentially solved intermediate variables $\{\z_1, \dots,\z_T\}$.
The final reconstruction result $\x$ after progressive splitting could be obtained according to $\z_T$ as 
\begin{equation}\small\label{eq_final_x}
    \hat{\x} = \argmin_{\x}  {\mu_T}  \lVert \z_T - \F \x \rVert^2_2,
\end{equation}
which has a simple closed-form solution via directly taking the inverse Fourier transform as $\hat{\x} = \F^{-1} \z_T$.

Now the problem lies in the solution of each intermediate auxiliary measurement $\z_t$. 
The update process of variable $\z_t$ in~\eqref{eq_update_zt} can be reformulated by merging the squares as 
\begin{equation}\small\label{eq_merge_zt}
\begin{split}
    \z_t =& \argmin_{\z_t\in \mathrm{R}(A^{\herm}_t)}  \lVert U_t \z_t -  \bm{v}_t \rVert^2_2
    + \lambda_t \mathit{\Psi}_t(\z_t),\\
\text{where} \quad & U_t^\top U_t = \mu_{t-1}D_{t-1}^\top D_{t-1} + \mu_t I, \\ 
& U_t^\top \bm{v}_t  = \mu_{t-1}D_{t-1}^\top \z_{t-1}+\mu_t A_t \x.
\end{split}
\end{equation}
Note that $U_t^\top U_t = \text{diag}(\bm{u}_t), \ \bm{u}_t\in \{\mu_{t-1}, \mu_{t-1} + \mu_{t}\}^n$ is a diagonal matrix which is easily invertible.
Thus it is equivalent to solve
\begin{equation}\small\label{eq_invert_zt}
    \z_t = \argmin_{\z_t\in \mathrm{R}(A^{\herm}_t)}  \lVert  \z_t -  U_t^{-1} \bm{v}_t \rVert^2_{U_t^\top U_t}
    + \lambda_t \mathit{\Psi}_t(\z_t),\\
\end{equation}
where $\lVert  \cdot \rVert^2_{U_t^\top U_t}$ is the weighted Euclidean norm~\cite{buades2005non} according to the diagonal of $U_t^\top U_t$.
From a Bayesian perspective, the optimization in~\eqref{eq_invert_zt} corresponds to a denoising problem with noisy input $U_t^{-1} \bm{v}_t$, which can be implicitly solved with neural networks.
Thus, the subproblem in~\eqref{eq_update_zt} can be solved in three steps: 
1) Calculate $U_t^{-1} \bm{v}_t$ which essentially performs data consistency on the results from the previous reconstruction $\z_{t-1}$;
2) Obtain a denoising result via a neural network $N_{\theta_t}$;
3) Projection onto the range space of the decomposed degradation $\mathrm{R}(A^{\herm}_t)$.

Finally, the overall optimization problem in~\eqref{eq_prob_P1} can be solved by breaking down the original severe degradation $A$ and reconstructing each decomposed degradation $D_t$ accordingly.
The whole iterative paradigm is described in Algorithm~\ref{alg1}, dubbed \textit{\textbf{progressive divide-and-conquer}}.
The final reconstruction result can be obtained by unfolding the proposed iterations into an end-to-end trainable network.

\subsection{Learning to Decompose}
So far, we have derived the proposed progressive divide-and-conquer framework based on the subsampling decomposition, which provides an indication of designing the data flow of our unfolded network.
An important question that arises is: \textit{how to obtain such a set of decomposed degradation $\{A_t\}_{t=1}^T$ based on the original degradation process $A$?}
Ideally, a straightforward solution is to jointly optimize such decomposition together in~${P_1}$, that is
\begin{equation}\small\label{eq_np-hard}
   \min_{\x, \{A_t\}_{t=1}^T}   \lVert \y - A_0 \x \rVert^2_2  + \sum_{t=1}^T \lambda_t \mathit{\Psi}_t(A_t \x).
\end{equation}
Unfortunately, the exact determination of such a decomposition set in~\eqref{eq_np-hard} proves to be an NP-hard problem~\cite{tillmann2013computational}. 
A remedy to the above problem is to gradually predict the decomposed degradation along with the PDAC iteration.
Specifically, a degradation predictor is introduced to estimate a sampling matrix $\bar{D}_t$ that characterizes the decomposed degradation $A_t$ based on the current reconstruction $\tilde{\z}_t$.
The degradation step in each iteration can be interpreted as retaining well-reconstructed frequency in the $k$-space while discarding the inaccurate information.
Following Definition~\ref{definition}, the severity of the decomposed degradation becomes weaker as $t$ increases. 
Accordingly, $\bar{D}_t$ tends to preserve more restored frequency as the reconstruction becomes more reliable with iterations until $\bar{D}_T=I$.

\noindent \textbf{Predicting subsampling mask.}
In the context of MRI, the sampling matrix refers to a subsampling mask $M$ in the $k$-space.
Specifically, we consider the Cartesian mask (\eg as shown in Figure~\ref{fig_framewwork}), which is the most commonly used in accelerated MRI.
The Cartesian mask corresponds to selecting frequency columns in the $k$-space. Thus, it can be represented by a binary vector $\m$ indicating the location of selected frequency sub-bands, \ie $M = \m \mathbf{1}^\top$, with $\mathbf{1}$ represents an all-one vector.
Following Definition~\ref{definition}, the subsampling mask should fulfill two properties:
1) the number of the sampled frequency columns should increase with iteration until a full mask;
2) the location of the sampled frequency columns should encompass the previous one.

\begin{algorithm}[t]
\setstretch{1.3}
\caption{Progressive Divide-And-Conquer (PDAC)} 
\label{alg1}
\begin{algorithmic}
\REQUIRE $\y$, $\mu_t$, $\N_{\theta_t}$, $A_t = \bar{D}_t \F$ defined in Section~\ref{subsec_decompose}.\\
\textbf{Initialization:}  $\z_0 = \y$, $\x = \F^{-1} \y$
\FOR{$t = 1,\dots ,T$}   
\STATE $U_t^\top U_t = \mu_{t-1}D_{t-1}^\top D_{t-1} + \mu_t I$
\STATE $U_t^\top \bm{v}_t = \mu_{t-1}D_{t-1}^\top \z_{t-1}+\mu_t A_t \x$
\STATE $\z_t' = U_t^{-1} \bm{v}_t \qquad \ \ \qquad \qquad$ \textit{\textcolor{gray}{\# data consistency}}
\STATE $\tilde{\z_t} = N_{\theta_t}(\z_t') \qquad  \qquad \qquad$ \textit{\textcolor{gray}{\# reconstruction}}
\STATE $\z_t = \bar{D}_t \tilde{\z_t}\qquad \ \ \ \ \  \qquad \qquad$ \textit{\textcolor{gray}{\# degradation}}
\ENDFOR
\STATE $\hat{\x} = \F^{-1} \z_T$
\STATE $\textbf{return} \;\hat{\x}$
\end{algorithmic}
\end{algorithm}

Consequently, we propose to predict subsampling mask $\m_t$ conditioned on the current reconstruction $\tilde{\z}_t$ and previous mask $\m_{t-1}$.
To make it feasible for network learning, we pre-define a budget $b_t = \lVert \m_t \rVert_0$ that restricts the number of sampled frequency columns in each mask $\m_t$, allowing the network to learn the frequency column locations within $\m_t$.
In detail, a degradation predictor $\mathcal{P}_{\theta_t}$ is introduced to estimate a probability $\p_t$ indicating the location of frequency columns to sample, \ie $\p_t = \mathcal{P}_{\theta_t}(\tilde{\z}_t)$.
Thus, the $\m_{t}$ can be obtained via adding extra frequency columns, selected from the $b_t - b_{t-1}$ largest probability of $\p_t$, on the previous mask $\m_{t-1}$. For each index $i$
\begin{equation}\small
\begin{split}
    \m_t^{[i]} = \m_{t-1}^{[i]} + \mathbb{I}(i \in \text{idx}) \quad \forall \ i,\\
\end{split}
\end{equation}
where idx is the indices of the $b_t - b_{t-1}$ largest values in $\p_t \odot (1-\m_{t-1})$ with $\odot$ denotes the Hadamard product.

\noindent \textbf{Decomposed degradation loss.}
Ideally, the predicted probability $\p_t$ shall indicate the confidence on the corresponding frequency columns in $\tilde{\z}_t$.
We define such confidence concerning the normalized reconstruction error in each frequency column.
In detail, given the $k$-space ground truth $Y_{gt}$, the normalized reconstruction error $\e_t$ is given by 
\begin{equation}\small
    \e_t =  \frac{ \mathbf{1}^\top \tilde{Z}_t - \mathbf{1}^\top Y_{gt}}{ \mathbf{1}^\top Y_{gt} }.
\end{equation}
Then we convert the range of $\e_t$ into $[0, 1]$ as a reference of probability $\p_t$, \ie $\hat{\e}_t = 2 \times \text{Sigmoid}(|\e_t|) - 1$, where $|\e_t|$ denotes element-wise absolute value of $\e_t$.
Building upon this, we introduce a decomposed degradation loss between $\p_t$ and $1-\hat{\e}_t$ using $\ell_1$-norm as:
\begin{equation}\small
    \mathcal{L}_{prob} = \sum_t^T \lVert \m_t \odot (\p_t - (1- \hat{\e}_t) )\rVert_1.
\end{equation}

\noindent \textbf{Degradation severity conditioning.}
During each iteration of PDAC, the network aims to reconstruct an intermediate measurement from a specific degradation severity, \ie $\z_{t}\in \mathrm{R}(A^{\herm}_{t})$. 
Thus we further adopt a severity conditioning module $E_{\theta_t}$ to guarantee awareness of the degradation pattern in the $k$-space.
Rather than directly conditioning on the binary mask vector $\m_t$, we propose to embed the masked confidence $\m_t \odot \p_t$ to further incorporate the estimated probability on each sampled frequency column.
Motivated by~\cite{peebles2023scalable}, we conduct an adaptive layer norm (adaLN) for inserting the information of the degradation severity.

\subsection{Training Objectives}
The iteration of PDAC is unfolded into a trainable network and optimized in an end-to-end manner.
To optimize the final reconstruction result, following the previous works~\cite{guo2021over, guo2023reconformer, yiasemis2022recurrent}, we adopt the $\ell_1$-norm pixel-wise loss $\mathcal{L}_{rec} = \lVert \hat{X} - X_{gt}\rVert_1$.
The total loss is defined as follows
\begin{equation}\small
    \mathcal{L}_{total} = \mathcal{L}_{rec} + \alpha \mathcal{L}_{prob},
\end{equation}
where $\alpha$ denotes the balancing parameter.
\begin{table}
    \centering
    \adjustbox{width=0.9\linewidth}{
    \begin{tabular}{c|cc}
    \toprule
    Setting & \multicolumn{2}{c}{$b_0, \dots, b_T \quad(T=8)$}\\
    \midrule
    Multi-coil 8$\times$ & \multicolumn{2}{c}{[48, 192, 288, 320, 336, 352, 364, 376, 384]}\\
    \midrule
    Single-coil 8$\times$ & \multicolumn{2}{c}{[40, 160, 240, 264, 280, 292, 304, 312, 320]}\\
    \midrule
    Single-coil 4$\times$ & \multicolumn{2}{c}{[80, 160, 240, 264, 280, 292, 304, 312, 320]}\\
    \bottomrule
    \end{tabular}
    }
    \vspace{-2mm}
    \caption{The sampling budget schedule for each experiment.}
    \label{tab_budget_schedule}
    \vspace{-3mm}
\end{table}

\section{Experiments}
\subsection{Experimental Setups}
\noindent \textbf{Dataset.}
The experiments are conducted on the fastMRI dataset~\cite{zbontar2018fastMRI} and Stanford2D FSE dataset~\cite{stanford2d}.
For the fastMRI dataset, we run the experiments on both the single-coil and multi-coil knee data which consists of 1,172 complexed-value MRI scans with 973 scans for training and 199 scans for validation.
Each scan approximately provides 30-40 coronal cross-sectional knee slices.\\
\textit{Mutil-coil track}: Each slice in the MRI scans is sampled by a total $C=15$ coils and each slice is center cropped into a size of $384 \times 384$.
Following the previous setting~\cite{fabian2022humus, sriram2020end}, we conduct the reconstruction on $8 \times$ accelerated MRI, with a central fraction of $4\%$ on low-frequency sub-bands.\\
\textit{Single-coil track}: Following~\cite{guo2023reconformer, guo2021over}, all images are center cropped into a size of $320 \times 320$.
We reconstruct MR images at two acceleration factors of $4 \times$ and $8 \times$, with a central fraction of $8\%$ and $4\%$ on low-frequency, respectively.

We also run experiments of multi-coil reconstruction on the Stanford2D FSE dataset, which consists of 89 fully-sampled volumes.
Following~\cite{fabian2022humus}, we randomly sampled $80\%$ data for training and used the rest for validation.
We conduct the reconstruction on $8 \times$ accelerated MRI that central cropped into a size of $384 \times 384$.

\begin{table}
    \centering
    \adjustbox{width=0.95\linewidth}{
    \begin{tabular}{clccc}
    \toprule
         Dataset&Method&  PSNR$\uparrow$&  SSIM$\uparrow$& NMSE$\downarrow$\\
     \midrule
        \multirow{7}*{\rotatebox{90}{fastMRI knee}} &Zero-filled& 27.42 & 0.7000 & 0.0704\\
         &ESPIRiT~\cite{uecker2014espirit}& 28.20 & 0.5965 & 0.1442\\
         &U-Net~\cite{ronneberger2015u}& 34.18 & 0.8601 & 0.0151\\
         &E2E-VarNet~\cite{sriram2020end}& 36.81 & 0.8873 & 0.0090\\
         &HUMUSNet~\cite{fabian2022humus}& 36.84 & 0.8879 & 0.0092\\
         &HUMUSNet$^\star$~\cite{fabian2022humus}& 37.03 & \textbf{0.8933} & 0.0090\\
         &\cellcolor{gray!20}HUMUSNet+PDAC&\cellcolor{gray!20}\textbf{37.12} &\cellcolor{gray!20}0.8905 &\cellcolor{gray!20}\textbf{0.0085}\\ 
     \midrule
        \multirow{5}*{\rotatebox{90}{Stanford2D}}&Zero-filled& 29.24 & 0.7175 & 0.0948\\
         &U-Net~\cite{ronneberger2015u}& 33.77 & 0.8953 & 0.0333\\
         &E2E-VarNet~\cite{sriram2020end}& 36.48 & 0.9220 & 0.0172\\
         &HUMUSNet~\cite{fabian2022humus}& 35.43 & 0.9134 & 0.0219\\
         &\cellcolor{gray!20}HUMUSNet+PDAC& \cellcolor{gray!20}\textbf{36.77} & \cellcolor{gray!20}\textbf{0.9247} & \cellcolor{gray!20}\textbf{0.0166}\\ 
      \bottomrule
    \end{tabular}
    }
    \vspace{-2mm}
    \caption{The quantitative results of $8\times$ accelerated multi-coil MRI reconstruction using our proposed model and recent methods on the fastMRI knee and Stanford 2D datasets. ($^\star$) denotes the model takes extra adjacent slices in the input.}
    \vspace{-2mm}
    \label{tab_multicoil}
\end{table}

\begin{figure*} [t]
  \centering 
    \includegraphics[width=0.83\linewidth]{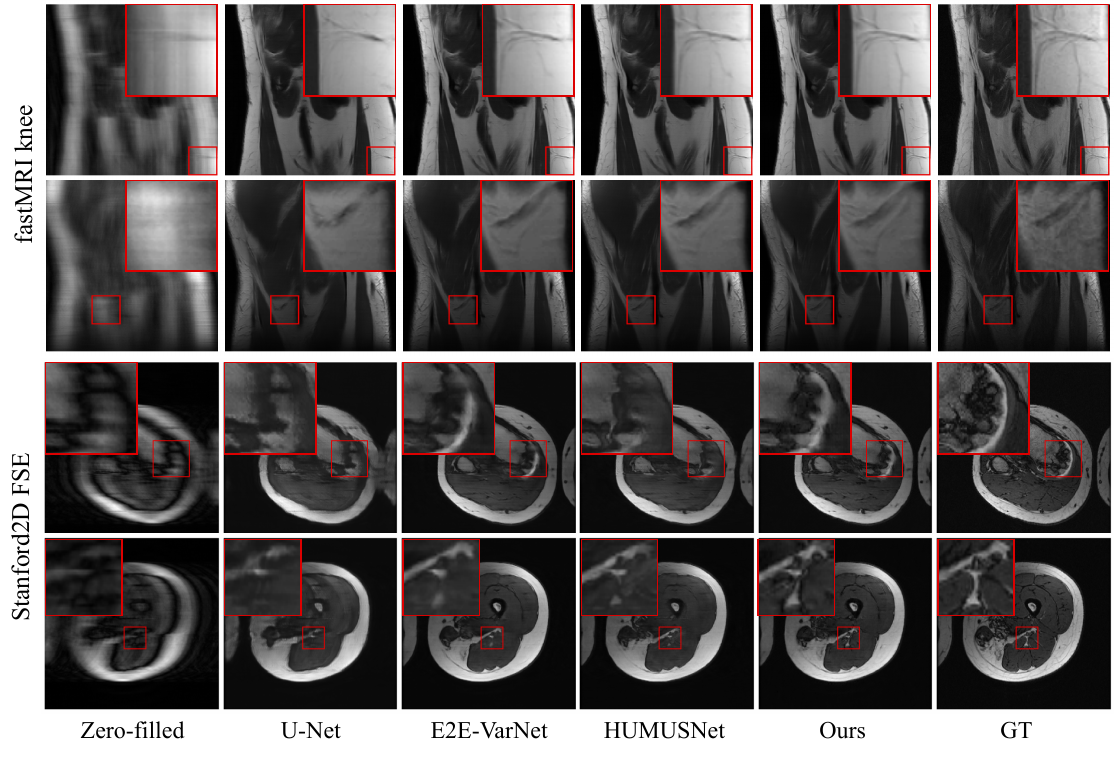}
    \vspace{-4.5mm}

\caption{Examples of multi-coil accelerated MRI reconstruction results of zero-filled input, U-Net~\cite{ronneberger2015u}, E2E-VarNet~\cite{sriram2020end}, HUMUSNet~\cite{fabian2022humus}, Ours and ground truth on the fastMRI~\cite{zbontar2018fastMRI} knee and Stanford2D FSE~\cite{stanford2d} datasets.
Please zoom in to see the details.}
    \label{fig_visual}
    \vspace{-2mm}
  \end{figure*} 

\noindent \textbf{Implementation details.}
The proposed method is implemented on PyTorch using NVIDIA RTX 4090 GPUs. 
We adopt the HUMUSNet~\cite{fabian2022humus} as the network backbone for our proposed PDAC framework.
The original HUMUSNet~\cite{fabian2022humus} takes a central slice together with two adjacent slices as the input for reconstruction, while we follow the common setting that only uses a single slice as the input for a fair comparison with the existing methods.
The detailed structures of degradation predictor $\mathcal{P}_{\theta_t}$ and severity conditioning module $E_{\theta_t}$ are provided in the supplementary material.
The network is trained in an end-to-end manner with $\mathcal{L}_{total}$ and optimized using AdamW~\cite{loshchilov2017decoupled}.
We train the network with a learning rate of $0.0001$ for $50$ and $40$ epochs on the multi-coil and single-coil settings, respectively, dropped by a factor of $10$ at the last $5$ epochs.
Following~\cite{fabian2022humus}, we set the total iteration step $T=8$ and we customize the column budget $\{b_t\}$ increasing in a coarse-to-fine manner, as detailed in Table~\ref{tab_budget_schedule}.
The weight for the probability loss $\alpha$ is set to $0.01$.

\subsection{Comparison with State-of-the-Art}
To evaluate the performance of our proposed PDAC strategy, we make a comparison with several state-of-the-art methods in both multi-coil and single-coil settings.
For the multi-coil setting, we adopt ESPIRiT~\cite{uecker2014espirit}, U-Net~\cite{ronneberger2015u}, E2E-VarNet~\cite{sriram2020end} and HUMUSNet~\cite{fabian2022humus} as the competing methods for $8\times$ accelerated MRI reconstruction.
Note that HUMUSNet$^\star$ indicates the results of the model using extra adjacent slices as the input for reconstruction.
For the single-coil setting, we compared the proposed methods with one classical method, \ie compressed sensing (CS)~\cite{tamir2016generalized} and eight learning-based methods, \ie U-Net~\cite{ronneberger2015u}, KIKI-Net~\cite{eo2018kiki}, Kiu-net~\cite{valanarasu2020kiu}, SwinIR~\cite{liang2021swinir}, D5C5~\cite{schlemper2017deep}, OUCR~\cite{guo2021over}, ReconFormer~\cite{guo2023reconformer} and HUMUSNet~\cite{fabian2022humus} on $4\times$ and $8\times$ accelerated reconstruction.

Tables~\ref{tab_multicoil} and~\ref{tab_singlecoil} show the quantitative results of the multi-coil and single-coil MRI reconstruction, respectively.
It is clear that our method can achieve superior performance consistently in both settings.
For the multi-coil reconstruction in Table~\ref{tab_multicoil}, the classical method ESPIRiT~\cite{uecker2014espirit} adopts a hand-crafted total variation regularizer that encourages sparsity, which is too restrictive for reconstructing details in the image.
Compared to the most recent method HUMUSNet~\cite{fabian2022humus}, the proposed PDAC method can improve the PSNR from $36.84$dB to $37.12$dB using the same network backbone.
Specifically, our method can even achieve a better PSNR with comparable SSIM results compared to HUMUSNet$^\star$ which uses extra information from adjacent MR slices in the input.
For the Stanford2D FSE dataset, our method significantly improves the PSNR for $1.34$dB compared to the HUMUSNet baseline.
A similar trend can also be observed in the single-coil setting, where the proposed model establishes new state-of-the-art in terms of both PSNR and SSIM in 8$\times$ reconstruction.

To further demonstrate the advantage of our method against other competing baselines, Figure~\ref{fig_visual} presents the visual examples of multi-coil MRI reconstruction results on fastMRI knee and Stanford2D FSE datasets.
 Our method excels in restoring MR images with enhanced details, particularly evident in the Stanford2D FSE dataset. 
 Notably, HUMUSNet~\cite{fabian2022humus} fails to recover tissue structures, as illustrated in the third row of Figure~\ref{fig_visual}, whereas our PDAC framework successfully restores the boundaries. 
 In the bottom row of Figure~\ref{fig_visual}, our method stands out as the sole approach that could restore a clear white triangle region, while other methods yield results plagued by blurry artifacts.

\begin{table}
    \centering
    \adjustbox{width=.95\linewidth}{
    \begin{tabular}{lcccc}
    \toprule
    & \multicolumn{2}{c}{$4\times$ acceleration} & \multicolumn{2}{c}{$8\times$ acceleration}\\
    \midrule
         Method&  PSNR$\uparrow$&  SSIM$\uparrow$& PSNR$\uparrow$&  SSIM$\uparrow$\\
     \midrule
            Zero-filled& 29.49 & 0.6541 & 26.84 & 0.5500\\
            CS~\cite{tamir2016generalized}& 29.54 & 0.5736 & 26.99 & 0.4870\\
            U-Net~\cite{ronneberger2015u}& 31.88 & 0.7142 & 29.78 & 0.6424\\
            KIKI-Net~\cite{eo2018kiki}& 31.87 & 0.7172 & 29.27 & 0.6355\\
            Kiu-net~\cite{valanarasu2020kiu}& 32.06 & 0.7228 & 29.86 & 0.6456\\
            SwinIR~\cite{liang2021swinir}& 32.14 & 0.7213 & 30.21 & 0.6537\\
            D5C5~\cite{schlemper2017deep}& 32.25 & 0.7256 & 29.65 & 0.6457\\
            OUCR~\cite{guo2021over}& 32.61 & 0.7354 & 30.59 & 0.6634\\
            ReconFormer~\cite{guo2023reconformer}& \textbf{32.73} & \textbf{0.7383} & 30.89 & 0.6697\\
            HUMUSNet~\cite{fabian2022humus}& 32.37 & 0.7221 & 31.04 & 0.6722\\
         \rowcolor{gray!20} HUMUSNet+PDAC& \textbf{32.73} & 0.7376 & \textbf{31.16} &\textbf{0.6739}\\ 
     \bottomrule
    \end{tabular}
    }
    \vspace{-2mm}
    \caption{The quantitative results of $4\times$ and $8\times$ accelerated single-coil MRI reconstruction using our proposed model and recent methods on the fastMRI knee dataset.}
    \vspace{-2mm}
    \label{tab_singlecoil}
\end{table}

\subsection{Ablation Study}
To investigate the effectiveness of each key component of the proposed method, we perform experiments on several model variants on the Stanford2D FSE dataset.

\noindent \textbf{The effect of degradation prediction.} 
We introduce adaptive learning for degradation decomposition in conjunction with the PDAC iterations.
However, one simple way to obtain such a decomposition is directly synthesizing each decomposed degradation based on the pre-defined budget $b_t$.
Thus we conduct an ablation study by comparing to the PDAC model with random decomposed degradation.
Specifically, the number of the frequency columns in $\bm{m}_t$ satisfies the same budget $b_t$ but the location of the frequency columns is randomly sampled from a uniform distribution.
However, the results indicate a significant degradation in performance, with a drop of approximately $0.31$dB in PSNR, compared to our proposed learning-to-decompose strategy, as shown in Table~\ref{tab_ablation}.
Our method predicts decomposed degradations based on the current reconstruction, akin to adaptively selecting well-recovered frequency components at each iteration. 
This progressive restoration of missing information in $k$-space unfolds gradually, from those that are easier to recover to more challenging ones.

\noindent \textbf{The effect of decomposed degradation loss.}
Following the aforementioned learning-to-decompose strategy, a probability $\bm{p}_t$ is estimated from a degradation predictor $\mathcal{P}_{\theta_t}$ to indicate the confidence of each reconstruction frequency column.
Such confidence is further regularized according to the normalized reconstruction error via the proposed decomposed degradation loss $\mathcal{L}_{prob}$. 
In order to illustrate the effectiveness of the decomposed degradation loss, we further conduct an experiment by training the PDAC model without $\mathcal{L}_{prob}$.
The results, detailed in Table~\ref{tab_ablation}, reveal a drop in PSNR from $36.77$dB to $36.48$dB, underscoring the contribution of the decomposed degradation loss to the learning of the degradation predictor.

\noindent \textbf{The effect of degradation severity conditioning.}
In the iteration of PDAC, the intermediate results are constrained to a specific subspace characterized by each decomposed sampling mask.
To ensure an awareness of the degradation pattern, we introduce a severity conditioning module $E_{\theta_t}$ that integrates the information of masked confidence, denoted as $\m_t \odot \p_t$. 
To investigate the effectiveness of this conditioning module, we further conduct experiments by removing $E_{\theta_t}$ in the PDAC iteration, leading to a deterioration in performance compared to our complete framework.

\begin{figure}
    \centering
    \begin{minipage}[c]{.43\linewidth} 
    \centering
    \includegraphics[width=0.95\linewidth]{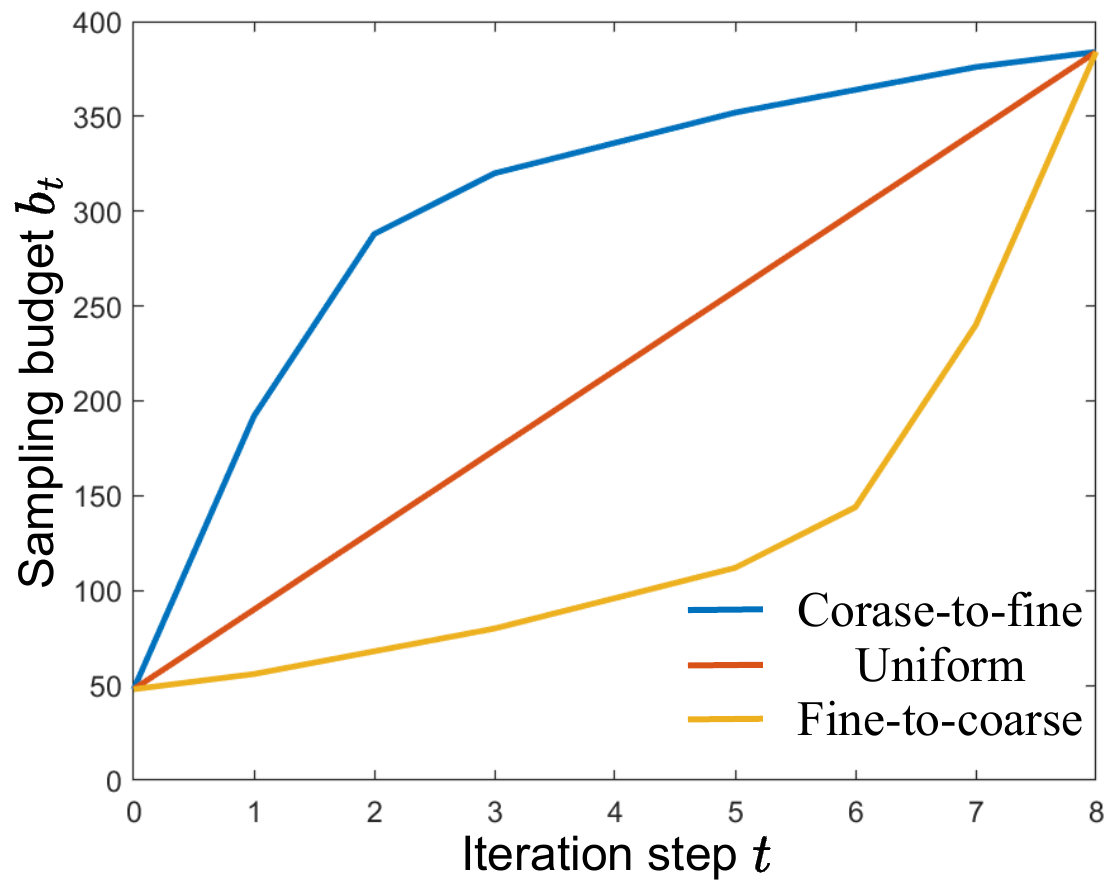}
    \vspace{-2mm}
    \end{minipage}
    \begin{minipage}[c]{.56\linewidth} 
    \centering 
            \adjustbox{width=.95\linewidth}{
            \begin{tabular}{lccc}
            \toprule
        
                  Sampling Schedule&  PSNR$\uparrow$&  SSIM$\uparrow$& NMSE$\downarrow$\\
             \midrule
                    Fine-to-coarse& 34.49 & 0.903 & 0.0469\\
                    Uniform& 36.66 & \textbf{0.925} & 0.0167\\
                    \rowcolor{gray!20}Coarse-to-fine & \textbf{36.77} & \textbf{0.925} & \textbf{0.0166}\\ 
             \bottomrule
            \end{tabular}
            }
            \vspace{-2mm}
            \vspace{-2mm}
            
    \end{minipage}
    \caption{Ablation study on different sampling budget schedules.}
    \vspace{-3mm}
    \label{fig_ablation_schedule}
\end{figure}

\noindent \textbf{The effect of sampling budget schedule.}
We conduct experiments on various schedules of the sampling budget ${b_t}$, including coarse-to-fine, uniform, and fine-to-coarse schedules. 
Figure~\ref{fig_ablation_schedule} shows that coarse-to-fine scheduling yields the best performance in our PDAC iteration.
This aligns with the design of PDAC, which aims to progressively recover the entire null space, moving from easier to more challenging aspects. 
The inherent suitability of coarse-to-fine scheduling for PDAC is evident, facilitating a larger step at the initial iteration and finer steps in subsequent iterations to reconstruct detailed information.

\begin{table}
    \centering
    \adjustbox{width=.95\linewidth}{
    \begin{tabular}{lccc}
    \toprule

         Model&  PSNR$\uparrow$&  SSIM$\uparrow$& NMSE$\downarrow$\\
     \midrule
            HUMUSNet baseline& 35.43 & 0.9134 & 0.0219\\
            PDAC w/o degradation prediction& 36.46 & 0.9233 & 0.0176\\
            PDAC w/o degradation loss $\mathcal{L}_{prob}$& 36.48 & 0.9238 & 0.0176 \\
            PDAC w/o severity conditioning& 36.66 & 0.9241 & 0.0168\\
         \rowcolor{gray!20} PDAC (Complete model)& \textbf{36.77} & \textbf{0.9247} & \textbf{0.0166}\\ 
     \bottomrule
    \end{tabular}
    }
    \vspace{-2mm}
    \caption{Ablation study to verify the effectiveness of each component in our method on the Stanford2D FSE dataset.}
    \vspace{-2mm}
    \label{tab_ablation}
\end{table}

\section{Conclusion}
In this paper, we propose a progressive divide-and-conquer framework for accelerated MRI.
We first decompose the subsampling process of the severe degradation in accelerated MRI into a series of moderate degradations. 
Following this, we provide a detailed derivation of the PDAC reconstruction process, which is subsequently unfolded into an end-to-end trainable network. 
Each iteration in PDAC is dedicated to recovering partial information within the null space, specifically addressing a particular moderate degradation based on the decomposition.
Furthermore, as part of the PDAC iteration, we incorporate the learning of degradation decomposition as an auxiliary task using a degradation predictor and a severity conditioning module. 
Finally, our extensive experimental results consistently demonstrate the superior performance of the proposed PDAC strategy in both single-coil and multi-coil MRI reconstruction on the fastMRI and Stanford2D FSE datasets.

{
    \small
    \bibliographystyle{ieeenat_fullname}
    \bibliography{main}
}


\end{document}